\documentclass[conference]{IEEEtran}
\IEEEoverridecommandlockouts
\usepackage{cite}
\usepackage{amsmath,amssymb,amsfonts}
\usepackage{graphicx}
\usepackage{textcomp}
\usepackage{xcolor}
\usepackage{xspace}
\usepackage{subcaption}
\usepackage{tikz}
\usepackage{tabularx}
\usepackage{framed}
\usepackage{booktabs}
\usepackage{multirow}
\usepackage{algorithm}
\usepackage{xurl}
\usepackage{hyperref}
\pdfmapfile{+fonts.map}
\usepackage[noend]{algpseudocode}
\newif\ifdraft
\drafttrue
\draftfalse

\ifdraft
\newcommand{\nojan}[1]{\textcolor{red}{{\sf (NS:} {\sl{#1})}}}
\newcommand{\anees}[1]{\textcolor{blue}{{\sf (AA:} {\sl{#1})}}}
\newcommand{\nges}[1]{\textcolor{orange}{{\sf (Ng:} {\sl{#1})}}}
\newcommand{\tk}[1]{\textcolor{red}{{\sf (TG:} {\sl{#1})}}}

\else
\newcommand{\nojan}[1]{}
\newcommand{\anees}[1]{}
\newcommand{\nges}[1]{}

\newcommand{\tk}[1]{}
\fi

\newcommand{\Prv}{$\mathcal{P}$\xspace}
\newcommand{\Vrf}{$\mathcal{V}$\xspace}
\newcommand{\Cir}{$\mathcal{C}$\xspace}
\newcommand{\sys}{\texttt{HashEmAll}\xspace}
\def\BibTeX{{\rm B\kern-.05em{\sc i\kern-.025em b}\kern-.08em
    T\kern-.1667em\lower.7ex\hbox{E}\kern-.125emX}}
    
\begin{document}

\title{Gotta Hash 'Em All! Accelerating Hash Functions for Zero-Knowledge Proof Applications}

\author{Nojan Sheybani$^{1}$\text{*}, Tengkai Gong$^1$\text{*}, Anees Ahmed$^{2}$, Nges Brian Njungle$^2$,  \\ Michel~Kinsy{$^2$},~Farinaz Koushanfar{$^1$} \\
\small{\text{*}Equal contribution} \\
$^1$University of California San Diego, $^2$Arizona State University\\
$^1$\tt\small{\{nsheyban, tegong, farinaz\}@ucsd.edu}, $^2$\tt\small{\{aahmed90, nnjungle, mkinsy\}@asu.edu} }

\maketitle
\begin{abstract}
Collision-resistant cryptographic hash functions (CRHs) are crucial for security, particularly for message authentication in Zero-knowledge Proof (ZKP) applications. However, traditional CRHs like SHA-2 or SHA-3, while optimized for CPUs, generate large circuits, rendering them inefficient in the ZK domain. Conversely, \textit{ZK-friendly hashes} are designed for circuit efficiency but struggle on conventional hardware, often orders of magnitude slower than standard hashes due to their reliance on expensive finite field arithmetic. To bridge this performance gap, we present \sys, a novel collection of FPGA-based realizations for three prominent ZK-friendly hashes: Griffin, Rescue-Prime, and Reinforced Concrete. Each offers distinct optimization profiles, with both area-optimized and latency-optimized variants available, allowing users to tailor hardware selection to specific application constraints regarding resource utilization and performance. 

Our extensive evaluation shows that latency-optimized \sys designs outperform CPU implementations by at least $10 \times$, with the leading design achieving a $23 \times$ speedup. These gains are coupled with lower power consumption and compatibility with accessible FPGAs. 
Importantly, the highly parallel and pipelined architecture of \sys enables significantly better practical scaling than CPU-based approaches towards building real-world ZKP applications, such as data commitments with Merkle Trees, by mitigating the hashing bottleneck for large trees. This highlights the suitability of \sys for real-world ZKP applications involving large-scale data authentication. We also highlight the ability to translate the \sys methodology to various ZK-friendly hash functions and different field sizes.
\end{abstract}


\section{Introduction}

The rapid advancement of digital technologies has brought forth unprecedented challenges in data privacy and security. As a result, privacy-preserving computation techniques have gained significant attention, enabling computations on encrypted data while maintaining the confidentiality of the underlying information. Among these techniques, zero-knowledge proofs (ZKPs) have emerged as a powerful cryptographic primitive, allowing users to prove the validity of statements about their private data without revealing any additional information. ZKPs have found applications in various domains, including authentication \cite{lu2008pseudo, liu2011zero}, data and IP ownership \cite{sheybani2023zkrownn}, and emerging learning paradigms \cite{ghodsi2023zprobe, liu2021zkcnn, weng2021mystique}. However, the practical deployment of ZKP systems faces significant challenges due to their high computational overhead, necessitating careful co-design of software and hardware to achieve efficient and scalable implementations.

Recent research efforts have focused on the development of hardware accelerators for ZKPs \cite{ma2023gzkp, zhang2021pipezk}, aiming to alleviate the computational bottlenecks and improve the overall performance of ZKP proving systems. While these accelerators have shown promising results, they often target high-performance computing platforms that cannot be easily realized in consumer hardware.
Many of the existing works focus on building efficient proving systems, rather than accelerating the operations that are used in ZKP applications. This presents a unique challenge, as the design of application-focused accelerators must carefully balance performance, resource utilization, and energy efficiency, while still being universally compatible with existing ZK proving schemes and their accompanying accelerators.

To address this challenge, it is crucial to identify and optimize the key computational building blocks 
of ZKP applications. \nojan{When speaking of applications, it is important to highlight the setup of ZKPs. In general, computation is performed twice in a given ZKP application: once in the plaintext domain and once in the ZK domain. probably leave this out} In this work, we focus on a critical component in all zero-knowledge applications: collision-resistant, cryptographic hash functions. Collision-resistant, cryptographic hash (CRH) functions play a vital role in ensuring the security and integrity of ZKP protocols and applications. Within the context of proof generation, especially in protocols like zk-STARKs \cite{ben2018scalable}, CRH functions serve as the core cryptographic primitive. However, more importantly, CRH functions are important in applications and algorithms that utilize ZKPs, such as Merkle Trees \cite{petkus2024efficient} and recursive proofs \cite{kothapalli2022nova}. While standard NIST-standardized CRH functions have been extensively studied and optimized for general-purpose computing \cite{sklavos2003hardware}, their efficiency in the context of ZKP applications is limited by their complex algebraic structure and high multiplicative complexity. This has led to the development of ZK-friendly hash functions \cite{Ingonyama_2022}, which are designed to exhibit simplified algebraic structures and reduced multiplicative complexity, making them more suitable for ZKP computations.


ZK-friendly hash functions are specifically designed to achieve optimal performance within zero-knowledge proof (ZKP) systems, where traditional cryptographic hash functions like SHA-2 and SHA-3 incur substantial computational overhead. Despite their circuit-friendly structure, these specialized hashes still perform significantly slower in plaintext operations, primarily due to the complex algebraic structures and arithmetic operations over large finite fields intrinsic to ZKPs. These operations typically require computations on integers at least 128 bits wide, often exceeding hardware capabilities of resource-constrained devices such as lower-end FPGAs, which generally support much narrower arithmetic operations. To bridge this performance gap and enhance the practical deployment of ZK-friendly hashes, dedicated hardware acceleration becomes essential, facilitating their adoption in various cryptographic applications.


In this work, we present \sys, our novel, open-source library of highly-optimized ZK-friendly hash functions built on reconfigurable hardware. This work focuses on three unique, state-of-the-art ZK-friendly hash functions: Rescue-Prime \cite{szepieniec2020rescue}, Griffin \cite{grassi2023horst}, and Reinforced Concrete \cite{grassi2022reinforced}.
Each of these hash functions takes a different approach towards achieving high performance in the plaintext domain, however, they still are not practical hashing solutions.
\sys aims to be the definitive solution for hashing in ZKP applications. Our solution maintains the high performance of the chosen hash functions in the ZK domain while providing state-of-the-art plaintext performance. To the best of our knowledge, \sys is the first work to perform ZK-friendly hashing with runtime in the same order of magnitude as the non-ZK-friendly hashing schemes, such as SHA-3. The open-source nature of our work allows for our designs to be easily integrated into real-world applications.

In short, the contributions of this work are as follows:
\begin{itemize}
    \item We present \sys, a collection of novel, state-of-the-art FPGA implementations of three notable ZK-friendly hash functions: Rescue-Prime, Griffin, and Reinforced Concrete. \sys provides designs to optimize for both resource utilization and latency independently, so that users may choose a design that best fits their constraints.\nojan{\sys balances the tradeoff between resource utilization and throughput by providing several designs for each hash function}
    \item \sys's open-source repository\footnote{\url{https://github.com/ACES-STAM/HashEmAll}} contains highly-optimized modules that are used to build the ZK-friendly hash functions over a finite field. In this work, we use a finite field compatible with the BN254 elliptic curve, which has been used heavily in practice. Due to the thorough and modular design of \sys, our presented optimized designs can be used to accelerate operations for all applicable ZKP applications.
    \item \sys achieves up to 23$\times$ speedup for hashing when compared to a state-of-the-art Rust implementation of the hash functions evaluated on a powerful CPU. This performance is achieved through deployment of our solution on reconfigurable hardware. \sys enables fast hashing on consumer hardware, spurring a paradigm shift in the usability and applicability of ZK-friendly hash functions in real-world ZKP applications.
\end{itemize}
\section{Preliminaries}

\subsection{Zero-Knowledge Proofs}
Zero-Knowledge Proofs (ZKPs) are cryptographic constructs that enable a prover, denoted as \Prv, to demonstrate to a verifier, indicated as \Vrf, that they possess knowledge of a secret value $w$, without disclosing any information about $w$. In ZKP literature, computations are frequently described using a circuit \Cir, conceptually viewed as a function processing public and private inputs to produce a public output. Formally, \Prv constructs a proof $\pi$ affirming their knowledge of a secret $w$ such that $C(x; w) = y$, with $x$ and $y$ being the public inputs and outputs respectively. 

Prior to generating a proof in ZKP frameworks, the circuit \Cir undergoes \textit{arithmetization}, transforming the computation using efficient mathematical constructs like polynomials. Although many ZKP schemes exist, such as zk-SNARKs \cite{nitulescu2020zk}, zk-STARKs \cite{ben2018scalable}, and MPCitH \cite{delpech2021limbo}, all require some form of arithmetization for an efficient arithmetic representation of \Cir. The intermediate representations when arithmetizing \Cir (e.g. R1CS constraints in zk-SNARKS \cite{golovnev2021brakedown}), typically indicate the complexity of \Cir, directly impacting the proof generation runtime of \Prv. This metric is generally referred to as \textit{constraints}.

One of the core challenges that contribute to the large computational overhead of ZKP 
applications is the underlying arithmetic structures. Modern ZKP implementations typically require operations to be done over large prime finite fields (e.g. 254-bit fields) to ensure certain security guarantees; however, consumer hardware, such as standard FPGAs and CPUs, typically only support operations up to 64 bits. Due to this, the efficiency of ZK operations is determined by metrics such as multiplicative complexity and algebraic structure rather than traditional metrics. This has spurred researchers to focus on designing ZK-optimized variants of standard primitives, which focus on efficiency in the ZK domain. One of the most prominent research areas in this research focus is the development of ZK-friendly hash functions.

\subsection{ZK-Friendly Hash Functions}
\nojan{edit} 
 
Collision-resistant hash functions are extensively used in ZK applications in both prove generation and verification phases. The objective of ZK-friendly hash functions is to execute cryptographically secure collision-resistant hashing, while necessitating a minimal number of constraints since standard hash functions like SHA-3, perform poorly in these settings. ZK-friendly hash functions are frequently utilized in zk-SNARK and zk-STARK applications, such as recursive proofs \cite{kothapalli2022nova}, membership proofs \cite{deng2023zktree}, and data ownership with Merkle Trees \cite{jing2021review}. 
There has been a notable increase in research that presents new practical ZK-friendly hash functions, such as Monolith \cite{grassi2023hash} and Poseidon \cite{grassi2021poseidon}.
The overarching goal of research in this field is to produce candidates for hashing in the ZK domain. Each proposed candidate aims to provide new optimizations, such as minimizing prover runtime, proof size, or verification runtime. All prominent ZK-friendly hashes can be categorized as one of the three following classes \cite{taceo2023hashes} \nojan{taceo article}:

\begin{itemize}
    \item \textbf{Low-degree.} These hash functions utilize low-degree round functions, typically in the form $y=x^d$ for a field-dependent integer $d$. This simplicity leads to the requirement of multiple rounds to achieve an acceptable level of security. The tenured hash functions, such as MiMC \cite{albrecht2016mimc} and Poseidon \cite{grassi2021poseidon}, fall under this class. While these hash functions are fast in the plaintext domain, relative to other ZK-friendly hashes, they are often costly in the ZK domain.
    \item \textbf{Low-degree equivalence.} These hash functions utilize round functions in the form of $y=x^{\frac{1}{d}}$ for a field-dependent integer $d$. This is due to the fact that, in the ZK domain, this can be translated to proving knowledge of $x$ such that $y^d=x$. However, in the plaintext domain, $y=x^{\frac{1}{d}}$ must be directly evaluated. Hash functions in this class generally require fewer rounds to achieve adequate security and yield fewer constraints. This results in faster hashing in the ZK domain, when compared to low-degree approaches, but slower hashing in the plaintext domain. Two prominent hash functions in this class are Rescue-Prime \cite{szepieniec2020rescue} and Griffin \cite{grassi2023horst}.
    \item \textbf{Lookup Tables.} These hash functions only work with proof systems that support table lookups, such as Plookup \cite{gabizon2020plookup} and Arya \cite{bootle2018arya}, but offer great performance in both plaintext and ZK domain. Rather than using power maps as round functions, these hash functions break up field elements into smaller chunks and perform operations on these smaller chunks using lookup tables. These smaller results are then recomposed to build the final resulting field element. This class of hash functions yields the fastest plaintext performance, albeit still slower than traditional hash functions, while only requiring very few rounds to achieve acceptable security guarantees. 
    However, there exists a large amount of variance in the number of constraints necessary to represent these hash functions, as it depends on the field size that is being employed. 
    Reinforced Concrete \cite{grassi2022reinforced} and Monolith \cite{grassi2023hash} are the standout hash functions in this class.
\end{itemize}


Unlike standard cryptographic hash functions such as SHA-3, ZK-friendly hashes are specifically designed to be efficiently represented within arithmetic circuits used in ZK protocols. Standard hash functions, optimized for general-purpose computing architectures, result in prohibitively large arithmetic circuits, significantly increasing the computational cost and complexity of ZK proofs. In contrast, the specialized algebraic structures of ZK-friendly hashes, employing techniques like low-degree polynomials, inverse mappings, or lookup tables, enable a substantial reduction in the number of arithmetic constraints required, thus significantly improving proving efficiency and practicality in ZKP applications.

\subsection{Applications and Importance of ZK-friendly Hashes. }

\begin{figure}
    \centering
    \includegraphics[width=0.95\columnwidth]{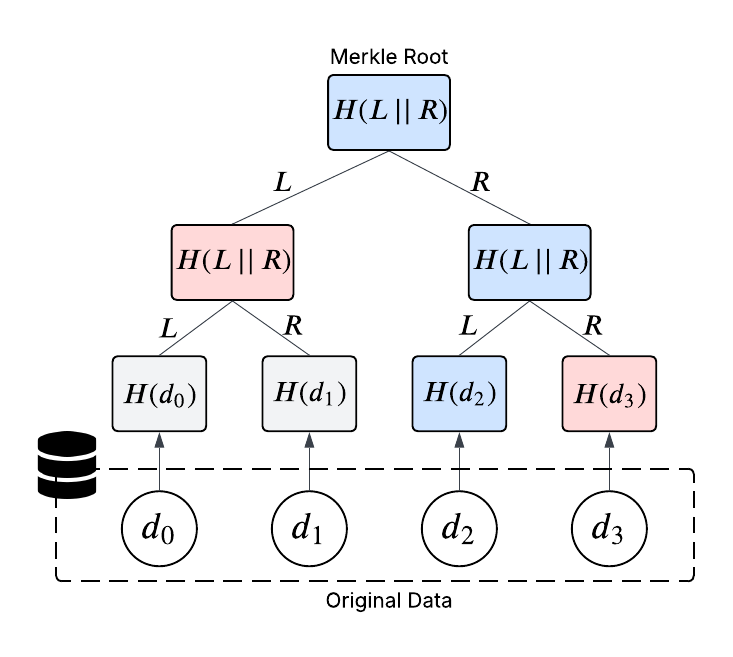}
    \caption{Example of a Merkle Tree constructed from a dataset with $4$ leaves (depth = $2$), illustrating the membership proof for data element $d_2$. Nodes on the Merkle path from $d_2$ to the root are highlighted in blue, while the required sibling nodes for the proof are highlighted in red.}
    \label{fig:merkle}
\end{figure}

ZK-friendly hash functions form the backbone of many zero-knowledge applications where hash-based structures and computations dominate the overall workload. One of the most prevalent examples is the use of Merkle Trees, which are integral to authenticated data structures in rollups, decentralized storage systems, and blockchain bridges. 
A Merkle Tree is a full binary tree whose leaves are the hashes of individual data blocks and whose internal nodes each store the hash of their two children, resulting in a single root hash that succinctly commits to all leaf values~\cite{jing2021review}. Representing a dataset with $N$ leaves results in a Merkle Tree with depth $\log N$. In zero-knowledge settings, the Merkle root serves as a public commitment, enabling a prover to demonstrate that a particular datum lies in the tree without revealing any other leaves. A membership proof, as illustrated in figure~\ref{fig:merkle}, consists of the target leaf together with the $O(\log N)$ sibling hashes along the path to the root, which the verifier, or the ZK circuit, re-hashes level by level to check consistency~\cite{deng2023zktree}. Because each proof requires $O(\log N)$ hash invocations, speeding up the underlying hash function directly translates into proportionate savings when generating or verifying large batches of Merkle proofs.
Our acceleration of ZK-friendly hash functions, particularly with highly parallel and pipelined architectures, directly reduces this cost, enabling much larger Merkle Trees, which can grow up to millions of leaves in current applications \cite{jing2021review}, to be handled in real time and making batch proofs viable on resource-constrained hardware.

Another key application is recursive proof composition, where one proof verifies the correctness of another. Recursive proofs use hash functions to commit to witness data, intermediate states, and prior proofs. Because these proofs are composed iteratively and must maintain tight runtime bounds, even modest inefficiencies in hashing can compound rapidly into significant performance overhead~\cite{kothapalli2022nova}.

Hash-intensive recursive proof systems are particularly important in blockchain applications. For instance, rollups, which are Layer-2 scaling solutions that batch transactions off-chain and submit aggregated data to the main blockchain, heavily rely on recursive proofs to securely and succinctly verify batched transactions~\cite{chen2022review}. Similarly, systems leveraging proof-carrying data or proof aggregation, such as private smart-contract state-transition proofs and zero-knowledge proofs for verifying machine-learning model integrity, demand extensive preprocessing of large hashed datasets~\cite{weng2021mystique,liu2021zkcnn}. By offloading hash evaluation to highly parallel accelerators, \sys significantly improves the practicality and scalability of these recursive proof-based applications, bridging the gap between ZK-efficient and plaintext-efficient hashing, and enabling higher throughput and wider deployability on mainstream hardware.

\section{\sys Hash Functions}

Our work focuses on hardware acceleration of three unique hash functions: Rescue-Prime, Griffin, and Reinforced Concrete. Each enacts a novel technique to balance ZK and plaintext performance. We do not consider low-degree round function hash functions, as previous works have provided sufficient research and hardware implementations \cite{githubGitHubDatenlordTRIDENT, ahmed2024amaze}. These hash functions excel in the ZK domain but struggle in plaintext. The goal of \sys is to make them practical candidates for faster ZKP applications by accelerating their plaintext operations on reconfigurable hardware.

We provide multiple hash options to allow users to choose the best function for their applications without performance penalties. Rescue-Prime and Griffin are compatible with all proof systems, with Rescue-Prime serving as the \textit{established} hash function and Griffin as the cutting-edge alternative. Reinforced Concrete offers excellent performance in lookup-supporting proof systems. When choosing between Rescue-Prime and Griffin, users should consider security analysis depth. Rescue-Prime has undergone a rigorous bounty program \cite{zkhashbounties2024}, making it ideal for those prioritizing thorough security analysis. Users comfortable with innovative technology can leverage Griffin's performance improvements. Applications using lookup-supporting proof systems will achieve optimal performance with Reinforced Concrete.

The chosen hash functions all operate over the prime field of the BN254 elliptic curve \cite{barreto2005pairing}, denoted as $\mathbb{F}_p$, due to its proven security guarantees. In this section, we highlight the permutations and round functions that serve as the unique aspects of each hash function. A ZK-friendly hash function is realized by using any of the presented permutations in the sponge framework. We denote the underlying permutation of any hash function H as $\texttt{H}_\pi$.

\nojan{add that all of them operate over finite field and use sponge construction so we can stop repeating ourselves}

\nojan{add low degree, low-degree equivalence, and lookup based items}

\nojan{justify where each one can be used (and why we chose them)}

\subsubsection{Rescue-Prime}

\begin{figure}[hbt]
    \centering
        \includegraphics[width=0.75\columnwidth]{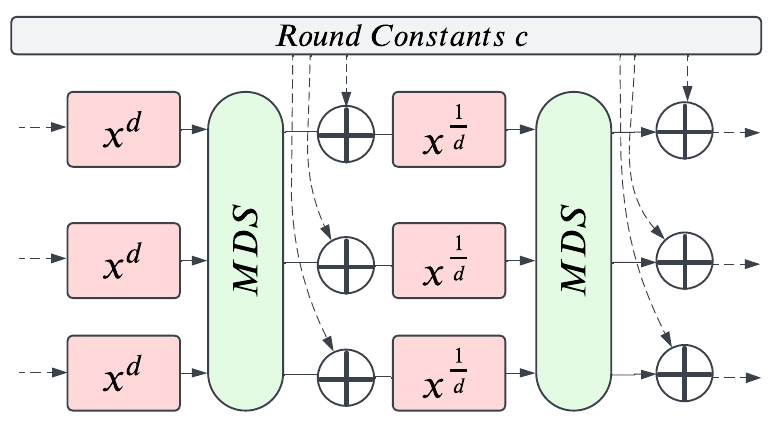}
    \caption{One round of the $\texttt{RescuePrime}_\pi$ permutation}
    \label{fig:rescue}
\end{figure}

Rescue-Prime \cite{szepieniec2020rescue} is the most tenured low-degree equivalence-based family of hash functions. We primarily consider this hash function due to its maturity and rigorous security analysis \cite{zkhashbounties2024}, serving as the option with established security for users with acceptable runtime in the ZK domain. \nojan{why hardware acceleration is necessary} Rescue-Prime utilizes a combination of power maps $x^{d}$ and $x^{1/d}$ as its round functions, where $d$ is chosen based on the specific field. 
The permutation \texttt{RescuePrime}$_\pi$, is repeated for $r$ rounds, each of which consists of the following operations for an input of state size $m$:
\begin{enumerate}
    \item S-box: Apply power map $x \mapsto x^d$ to all elements in state.
    \item MDS \& Constants: Multiply state with MDS matrix $M$ and add $m$ constants from list $c$.
    \item Inverse S-box: Apply inverse power map $x \mapsto x^{1/d}$ to all elements in state.
    \item MDS \& Constants: Multiply state with MDS matrix $M$ and add $m$ constants from list $c$.
\end{enumerate}
\noindent This is also illustrated in Figure \ref{fig:rescue}. We refer to \cite{sim2015lightweight} for the description of the MDS matrix, but note that it is simply part of a matrix-vector multiplication.
A visualization of $\texttt{RescuePrime}_\pi$ can be seen in figure \ref{fig:rescue}.

\subsubsection{Griffin}

\begin{figure}[htb]
    \centering
    \includegraphics[width=0.98\linewidth]{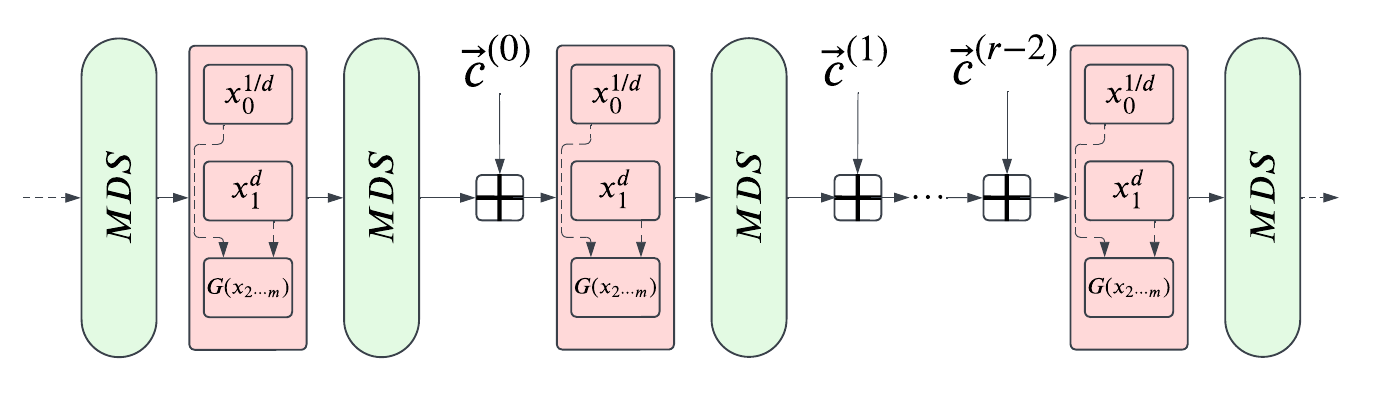}
    \caption{One round of the $\texttt{Griffin}_\pi$ permutation, where $\boxplus$ denotes element-wise addition}
    \label{fig:griffin}
\end{figure}

Griffin \cite{grassi2023horst} is a family of hash functions that is based on the Rescue-Prime hash function to achieve better plaintext performance while maintaining comparable ZK performance. Griffin uses a combination of power maps $x^d$ and $x^{1/d}$, along with a Horst-inspired construction in its round functions, where $d$ is chosen based on the specific field. The primary difference between Griffin and Rescue-Prime is Griffin's introduction of a quadratic function to its permutation function and sequential processing of state elements within the S-box, rather than Rescue-Prime's approach of applying the same function to every element in the state.

The permutation $\texttt{Griffin}_\pi$ is repeated for $r$ rounds \nojan{maybe add function to intro} with an input $x=(x_0, \cdots, x_m)$ of state size $m$, in which each round performs the following operations:
\begin{enumerate}
    \item Nonlinear layer: Rather than applying a power map or an inverse power map to each element in the state, $\texttt{Griffin}_\pi$'s nonlinear layer $S$ is defined as
    \begin{equation*}
        y_i = 
        \begin{cases}
            x_0^{1/d} & \text{if } i=0 \\
            x_1^{d} & \text{if } i=1 \\
            G(x_2, y_0, y_1) & \text{if } i=2 \\
            G(x_{i-1}, y_0, y_1) & otherwise
        \end{cases}
    \end{equation*}
    where $G(\cdot)$ is a quadratic function. For brevity, we refer to \cite{grassi2023horst} for more information on the construction of $G(\cdot)$. In terms of computation, $G(\cdot)$ is a trivial function to compute on hardware.
    \item Linear Layer: The state is multiplied by an MDS matrix $M$ and a vector of round constants $\vec{c}$ is added to the state.
\end{enumerate}
As demonstrated in Figure \ref{fig:griffin}, the first round of $\texttt{Griffin}_\pi$ has an extra step, in which the state is multiplied by an MDS matrix $M$. \nojan{better finish}




\subsubsection{Reinforced Concrete}

\begin{figure}[htb]
    \centering
    \includegraphics[width=0.98\linewidth]{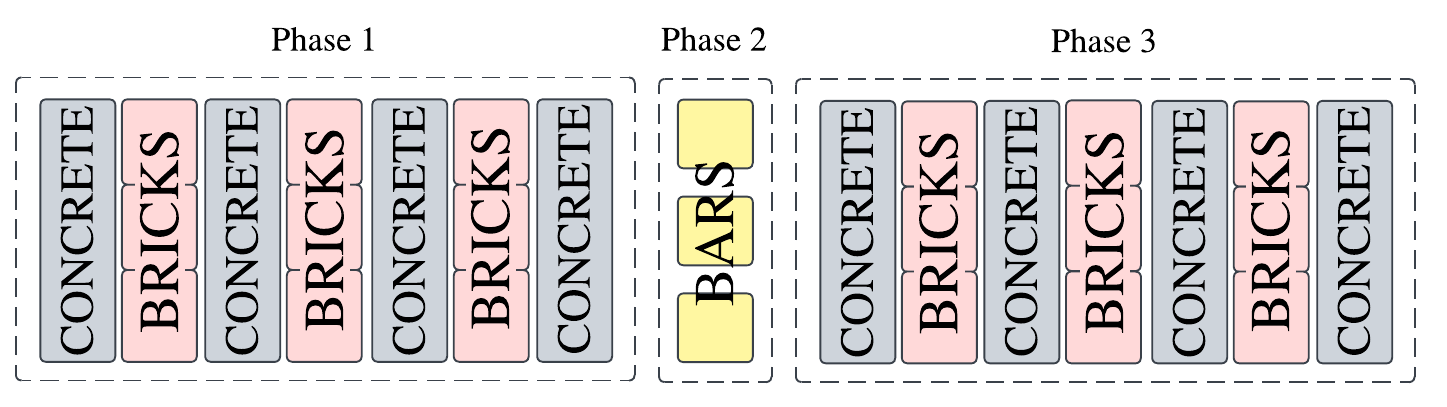}
    \caption{$\texttt{RC}_\pi$ permutation}
    \label{fig:rc}
\end{figure}

Reinforced Concrete \cite{grassi2022reinforced} (RC) is a lookup-based hash function that is optimized to achieve high performance when operating over large finite fields. One of the limitations of RC is that it can only achieve peak performance when operating over a proof system that supports lookup functions. However, in the right conditions, RC is one of the fastest ZK-friendly hash functions in the plaintext domain. RC utilizes a combination of lookup-based S-boxes and efficient linear layers to achieve this performance.

The permutation $\texttt{RC}_\pi$ does not operate over a variable amount of rounds. Instead, $\texttt{RC}_\pi$ is defined as a 7-round operation, in which the only supported state size $m$ is 3, meaning that all inputs are of form $(x_1, x_2, x_3)$. $\texttt{RC}_\pi$ consists of three components:
\begin{enumerate}
    \item \texttt{Concrete}: This is a linear operation in which the state is multiplied by an MDS matrix and a vector of round constants are added to the state.
    \item \texttt{Bricks}: This is a non-linear permutation of degree 5 that has similar computational complexity to the evaluation of $G(\cdot)$ in \texttt{Griffin}$_\pi$. Specifically,
    \begin{align*}
        &\texttt{Bricks}(x_1, x_2, x_3) \\
        &= (x_1^d,x_2(x_1^2+\alpha x_1 + \beta_1), x_3(x_2^2 + \alpha_2x_2 + \beta_2)),
    \end{align*}
where $\alpha_1, \beta_1, \alpha_2,\beta_2 \in \mathbb{F}_p$. \nojan{FIX F FOR EVERYTHING}
    \item \texttt{Bars}: This function processes states elements independently:
    \begin{equation*}
        \texttt{Bars}(x_1, x_2, x_3) = (\texttt{Bar}(x_1), \texttt{Bar}(x_2), \texttt{Bar}(x_3)),
    \end{equation*}
    where \texttt{Bar}$(\cdot)$ consists of the functions \texttt{Decomp}, \texttt{Sbox}, and \texttt{Comp}.
    \texttt{Decomp} breaks down the element into smaller chunks, decomposing each chunk into an element over a unique subfield $\mathbb{F}_{s_i}$. \texttt{Sbox} acts as a permutation function for the small chunks, which typically utilizes a lookup table for operation. \texttt{Comp} recomposes these small chunks into elements of the original finite field $\mathbb{F}_p$. This is a condensed description of this operation for brevity, but the overarching idea of the \texttt{Bars} operation is to independently process each element in the state with low latency, which is achieved by breaking down the elements into smaller chunks and applying \texttt{Sbox} on these smaller chunks. For more details, including the mathematical descriptions of these functions, we refer to \cite{grassi2022reinforced}.
\end{enumerate}

The structure of $\texttt{RC}_\pi$ is illustrated in \ref{fig:rc}. As can be seen, the first and third phases of the permutation only consist of \texttt{Concrete} and \texttt{Bricks}. The \texttt{Bars} function, which has the highest computationally complexity, is only called once, which is how RC is able to achieve relatively fast runtime in the plaintext domain.
Reinforced Concrete's performance varies depending on the specific field, but it generally achieves a 2-9x overhead compared to SHA-256 in native computations, while significantly outperforming previous ZK-friendly hash functions in proving systems \cite{grassi2022reinforced}.








\section{Related Works}

There are two primary threads of research focusing on hardware acceleration for ZKPs: ZKP applications and proof generation. Proof generation acceleration, involving operations like NTT and MSM, is well-studied. PipeZK~\cite{zhang2021pipezk} and cuZK~\cite{lu2023cuzk} accelerate proof generation on ASIC and GPU, respectively. NoCap~\cite{samardzicaccelerating}, a vector processor, achieves 41$\times$ speedup compared to PipeZK. While valuable, these works are tangential to our research, as \sys aims to accelerate ZKP applications and can be used alongside these accelerators for optimal performance.

Hardware acceleration for ZKP applications has primarily focused on ZK-friendly hash functions. Notable works include TRIDENT~\cite{githubGitHubDatenlordTRIDENT} and AMAZE~\cite{ahmed2024amaze}, which accelerate Poseidon and MiMC hash functions, respectively. These low-degree-based hash functions are popular in current ZKP applications due to their simplicity and security analysis~\cite{zkhashbounties2024}. However, they are natively fast in plaintext but slow in the ZK domain.

\sys focuses on hash functions based on low-degree equivalence (Rescue-Prime, Griffin) or lookups (Reinforced Concrete). Compared to low-degree-based functions, these achieve better performance in the ZK domain and, with \sys, comparable or better performance in plaintext. This approach allows for a more balanced implementation across both domains. \sys also enables users to choose from these three hash functions based on their application constraints.
\section{Methodology}
\sys achieves state-of-the-art performance for the chosen ZK-friendly hash functions while employing a modular design approach. We take this approach to enable reusability of the modules between hash functions. These modules are optimized for latency and resource utilization and, although designed for Griffin, Rescue-Prime, and Reinforced Concrete, can be utilized in any ZK-friendly hash functions with similar operations. \sys employs an open-source FPGA library \cite{ahmed2024amaze} with optimized modules for  \anees{``AMAZE open-source FPGA library'' reads a bit like advertisement to me, since an average reader would not refer to it like that by reading our paper. I feel like it might also give away our identity during blind review.} \tk{shall we just use [34] then?}to enable accelerated computation of the necessary arithmetic over the prime field of the BN254 elliptic curve \cite{barreto2005pairing}. We will denote the prime field as $\mathbb{F}_p$. In this section, we outline the design and motivation of our novel modules that enable \sys's state-of-the-art realizations of the chosen ZK-friendly hash functions and then outline how these modules are used to accelerate each hash function. We refer the reader to \cite{ahmed2024amaze} for the details on optimized and pipelined implementation of modular multiplication and other arithmetic operations, which form the basis for the higher-level routines and operations that constitute the proposed hardware of the three hashes in this work.
\nojan{We exclude descriptions of trivial operations and the operations that were optimized by the AMAZE library - idk something like this} \anees{Alternatively: }
\nojan{Performing operations over this prime field allows us to achieve \nojan{X} bits of security in \sys.} \anees{Typically, hashes aim to have roughly 128 bits of security. E.g. Refer to the sponge security formula in section ``Truncation and generic security`` in the Rescue-XLIX paper}

\subsection{Fast Divisions with Lookup Tables}
\tk{This should go before talking about reconfigurable multiplier}
Division is natively slow and resource-intensive in hardware, however, it is a core operation in \texttt{RC}$_\pi$'s \texttt{Decomp} function. Division is the primary way we are able to reduce an input to the \texttt{Bar} function to an element in a unique subfield $\mathbb{F}_{s_i}$  Luckily, we observed that in many cases, the divisor, in this case $s_i$, is predefined, which allows us to employ a different approach to division. Rather than doing any division, we precompute the reciprocal of the divisor, which is $1/s_i$. This value is scaled by a factor $D$, which is a power of two, to be compatible with fast division via bit shifting, and also double the size of the input bit (meaning $D$ is 508 bits in our BN254 setting). This value, $D/s_i$ is then stored in a lookup table (LUT). The calculation is then converted to $Result = (x\times \frac{D}{s\_i}) >> 508$, in which  $\frac{D}{s\_i}$ is a simple lookup.

\subsection{Reconfigurable Modular Multiplier}
\nojan{talk about two multiplication modes and add figure}
\tk{Used in reinforced concrete only}
There are 3 arithmetic operations in RC - quadratic function evaluation, \texttt{Decomp}, and \texttt{Comp}. To enable this functionality, we build a reconfigurable modular multiplier that has different modes for handling these computations. For quadratic function evaluation, the modular multiplier is instantiated in the same way as it is in \texttt{Griffin}$_\pi$. We denote this mode as \texttt{MULT}.

For $\texttt{Decomp}$, the needed modular multiplier is one that can perform the lookup table approach for division that is described in the previous subsection. As the result of this operation exceeds 256-bit, this requires a multi-step approach to properly reduce the result of the multiplication. We denote this mode as \texttt{DECOMPOSE}.

For \texttt{Comp}, since the subfields are small, we can skip modular reduction in the first few steps of recomposing the elements of the state. Instead, we instantiate three multipliers to run in parallel to calculate the product and recompose the state element as an element in the original field $\mathbb{F}_p$. We denote this mode as \texttt{COMPOSE}.

\subsection{Fast Power Mapping}
One of the most important operations in hash functions that are based on low-degree equivalence is the power map, in which an input is raised to the power of some field element $d$. In Griffin and Rescue-Prime, both the power map $x \mapsto x^d$ and its inverse power map $x \mapsto x^\frac{1}{d}$ are utilized. This operation enables fast performance in the ZK domain but serves as the main bottleneck in plaintext realizations. For instance, in the case of hash functions over BN254, this amounts to raising a 254-bit field element with an exponent that itself is 254-bit wide. To accelerate this operation over $\mathbb{F}_p$, we use the square-and-multiply algorithm \cite{hui1994fast}, shown in Algorithm \ref{alg:GFp_exp_square_multiply}. This approach requires $\mathcal{O}(n)$ rounds of computations for a given bitwidth $n$ of the exponent. When optimizing for latency, this module can be instantiated with two modular multiplication units that are used in parallel. When optimizing for resource utilization, instantiating with one modular multiplication unit is sufficient.
\nojan{talk about hardware implementation} \tk{It might be clearer to use algorithm}
\begin{algorithm}[!h]
    \caption{Square-and-Multiply Algorithm for Modular Exponentiation}
    \label{alg:GFp_exp_square_multiply}
    \begin{algorithmic}[1]
        \Require $x \in \mathbb{F}_p$, $d_{\text{inv}} \in \mathbb{F}_p$, modulus $p$
        \Ensure $y = (x^{d_{\text{inv}}}) \bmod p$
        \State $result \gets 1$
        \State $base \gets x$
        \While {$d_{inv} > 0$}
            \State $result \gets $\textbf{if} $\Call{MSB}{d_\textit{inv}}$ \textbf{then} $result \times base$ \textbf{else} $result$ 
            \State $base \gets base \times base$ 
           \State $d_{inv} \gets d_{inv}>> 1$
           \EndWhile
           \State $y \gets result$
    \end{algorithmic}
\end{algorithm}

\subsection{Sponge Function}
\begin{figure}[htb]
    \centering
    \includegraphics[width=0.98\linewidth]{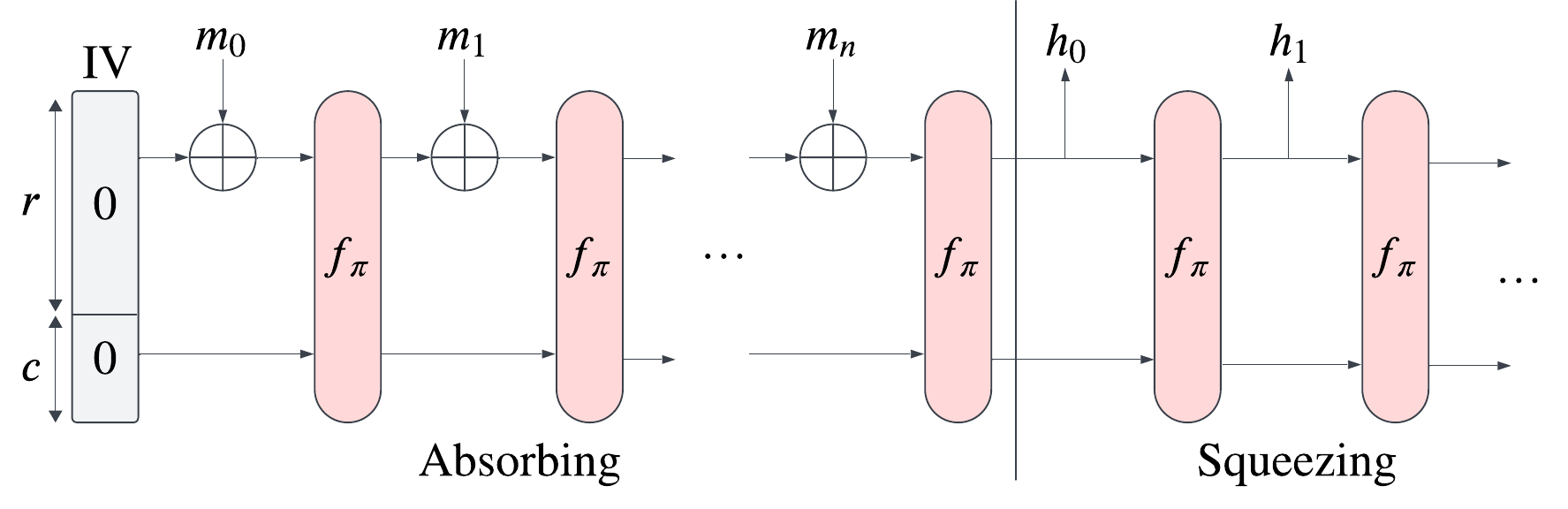}
    \caption{Sponge framework for permutation $f_\pi$. This framework absorbs the message $m$ and outputs the hash $h$. \anees{put caption here}}
    \label{fig:sponge}
\end{figure}

Sponge functions are cryptographic primitives used to process input data into an output of arbitrary length \cite{bertoni2007sponge}. A sponge function can be used to convert any fixed-length permutation into a collision and pre-image-resistant hash function, assuming that the permutation is also secure. They consist of two main phases: absorption and squeezing. During the absorption phase, the initial input data chunk is XORed into the initial state of the sponge's internal structure and then transformed by the permutation function. Then, the output of the permutation function, XORed with the next input data chunk is used as an input to the next permutation function. This operation is repeated until the input data has been fully read. In the squeezing phase, the state is transformed to produce output chunks until the desired output length is reached, resulting in the hashed value of the initial input. This process is shown in figure \ref{fig:sponge}. The hash functions that are accelerated in this work all operate in the sponge framework.

\subsection{Fast Hashing with \sys}

All of the hash functions can be instantiated in the sponge framework.
\sys employs a pipelined approach for maximizing hash throughput for each hash function. These hash functions were chosen due to their high performance in the ZK domain. In what follows, we highlight how the modules that we have developed enable high performance in the plaintext domain. For each hash function $H$, we present two designs: $H_A$, a design that aims to minimize area and resource utilization and $H_L$, a design that aims to maximize amortized latency and throughput. All of the permutations are built with a state size of 3, meaning that the permutation handles 3 elements at a time. For area-optimal designs, we instantiate a pipeline that can process a batch size of 13 inputs at once. For designs that optimize for latency, we support batch sizes in multiples of 13 by instantiating parallel pipelines. Although this works for all of our presented hash functions, we aim to have the latency-optimized solutions all utilize the same amount of resources roughly, to ensure fair comparison. The evaluation of our designs is done with this in mind. For one permutation, $\texttt{RescuePrime}_\pi$ and $\texttt{Griffin}_\pi$ both require 14 round function evaluations, while $\texttt{RC}_\pi$ only requires 1. \nojan{We aim to have the latency-optimized solutions all be around the same size. also talk about why we have batch size of 13} 

\nojan{All instantiated in the sponge hashing framework}
\nojan{These all already have fast ZK performance, so we want to focus on plaintext performance}
\nojan{We present latency-optimized and resource-optimized implementations}

\noindent\textbf{Rescue-Prime}
The core non-trivial arithmetic that is used in \texttt{RescuePrime}$_\pi$ is power mapping and inverse power mapping. These operations are accelerated with our square-and-multiply-based algorithm for fast power mapping. Design $\texttt{RescuePrime}_A$ minimizes resource utilization by only instantiating three modular multiplier units and pipelining their usage. Design \texttt{RescuePrime}$_L$ minimizes latency and maximizes throughput by instantiating six modular multiplier units. We use 6 modular multiplication units as the latency-optimized power mapping module requires 2 parallel multipliers to achieve low latency. Due to our state size being 3, and due the fact that \texttt{RescuePrime}$_\pi$ performs the power map across the whole state, \nojan{be consistent with numbers} we instantiate 6 modular multipliers in parallel to minimize the latency of applying the power map and the inverse power map.

\nojan{Primarily uses AMAZE's fast exponentiation and fast inverse power mapping. Linear Layer is trivial} \anees{exponentiation from amaze doesn't end up being useful here. so we're just using the pipelined modmul from amaze work, to build forward and inverse power maps using simple binary exponentiation method.}

\noindent\textbf{Griffin}
\sys's design of $\texttt{Griffin}_\pi$ is not so different from that of $\texttt{RescuePrime}_\pi$, as they both perform power mapping and inverse power mapping. However, as we mention before, $\texttt{Griffin}_\pi$ does not apply the power maps to the whole state, but rather only to the first two elements in the state. The third element evaluates a quadratic function that utilizes the previous elements in the state. This quadratic function is trivial to compute, especially if certain elements of it are precomputed. We design $\texttt{Griffin}_\pi$ with this precomputation in mind. Just like in \texttt{RescuePrime}$_L$, we instantiate the power mapping and inverse power mapping modules with two multipliers for the square-and-multiply-based exponentiation. However, one of the multipliers is sometimes idle (seen in line 4 of algorithm \ref{alg:GFp_exp_square_multiply}. During this idle phase, the multiplier is used to precompute parameters for the quadratic function, as well as to apply the power map $x \mapsto x^d$.

\renewcommand{\arraystretch}{1.1}
\begin{table*}[t]
    \centering\resizebox{\textwidth}{!}{
    \begin{tabular}{cccccccccccc}
        \toprule
        \multicolumn{2}{c}{\textbf{}} & \multicolumn{5}{c}{\textbf{Resource Utilization}} & \multicolumn{4}{c}{\textbf{Timing}} \\
        \cmidrule(lr){3-7}
        \cmidrule(lr){8-11}
         & Batch &  &  &  &  &  & Amortized &Amortized &  &  & \\
         Design \nojan{?} & size  & LUTs & FFs & LUTRAM & DSPs & Freq. (MHz) & Latency (\textmu s) & Throughput (kops/s) & Power (W)\\
        \midrule
         \texttt{Rescue-Prime}$_A$ & 13  & 128,519 (7.4\%) & 59,301 (2\%) & 5,968 & 1,761 (14.3\%) & 100.21\tk{change} & 56.04 &17.84 & 6.42\tk{(change} \\
         \texttt{Griffin}$_A$ & 13 & 76,538 (4\%)  &68,514 (1.98\%) & 665 & 1,174 (9.95\%) & 103.88 & 34.7 &28.8 & 7.125 \\
         \texttt{RC}$_A$ & 13 & 156,154 (9\%) & 84,921 (2\%)  & 2,821 & 2,016 (16\%) & 99.95 & 0.28 &3,571 & 10.425 \\     
         \hline
         \texttt{Rescue-Prime}$_L$ & 13 & 161644 (9\%) & 79219 (2\%) & 1353 & 3522 (29\%) & 100.13 & 36.5 &27.39 & 13.237 \\
         \texttt{Griffin}$_L$ & 39 & 224,949 (13\%) & 144,926 (4.2\%)  & 1995 & 3,522 (28.7\%) & 103.93 & 11.57 &  86.43 & 15.181  \\
         \texttt{RC}$_L$ & 26 & 316,049(18.28\% & 169,043(4.89) & 1,524  & 4,032(32.81\%) & 96.56 & 0.145 &6,896 & 22.324\tk{change} \\
         \bottomrule
    \end{tabular}}
    \caption{Evaluation of \sys on Virtex Ultrascale+ \nojan{board}}
    \label{tab:hashemall_benchmark}
\end{table*}
\tk{Consider deleting Cycles. Latency should be enough }

\renewcommand{\arraystretch}{1.2}
\begin{table}[htbp]
    \centering
    \resizebox{0.95\columnwidth}{!}{
    \begin{tabular}{cccccc}
        & \multicolumn{2}{c}{\multirow{2}{*}{\begin{tabular}[c]{@{}c@{}}\textbf{Amortized}\\\textbf{Throughput (kops/s)}\end{tabular}}} & \multicolumn{2}{c}{\multirow{2}{*}{\begin{tabular}[c]{@{}c@{}}\textbf{Amortized}\\\textbf{Latency ($\mu$s/hash)}\end{tabular}}} & \multirow{2}{*}{\textbf{Speedup}} \\
        \textbf{Hash} & & & & & \\
        \cline{2-6}
        & CPU & \sys & CPU & \sys & vs. CPU \\
        \hline
        \texttt{Rescue-Prime}$_A$ &  2.4 & 17.84\tk{X} & 415 & 56.04\tk{X} & 7.4$\times$ \\
        \texttt{Griffin}$_A$ & 8.695 & 28.8 & 115 & 34.7 & 3.31$\times$ \\
        \texttt{RC}$_A$ & 294.12 & 3,571 & 3.4 & 0.28 & 12.14$\times$  \\
        \hline
        \texttt{Rescue-Prime}$_L$ & 2.4 & 27.39 & 415 & 36.5 & 11.37$\times$  \\
        \texttt{Griffin}$_L$ & 8.695  & 86.43 & 115 & 11.57 & 9.94$\times$  \\
        \texttt{RC}$_L$ & 294.12 & 6896 & 3.4 & 0.145 & 23.44$\times$ \\
        \hline
    \end{tabular}}
    \caption{Performance Comparison between CPU and \sys}
    \label{tab:hash_performance}
\end{table}

\noindent\textbf{Reinforced Concrete}
\sys utilizes many of the optimized modules to achieve fast performance for $\texttt{RC}_\pi$. The core non-trivial operations in $\texttt{RC}_\pi$ are quadratic function evaluation, field element decomposition with \texttt{Decomp}, and field element composition with $\texttt{Comp}$. These are all accelerated with our novel reconfigurable modular multiplier, which contains submodules for fast division and quadratic function evaluation. 

For our design that minimizes resource utilization $\texttt{RC}_A$, we only instantiate three reconfigurable modular multiplication units. The reconfigurable units are optimized to switch between modes based on the computation that is presented to them, such as switching to \texttt{DECOMP} mode when entering the \texttt{BARS} function.
The design that minimizes latency and maximizes throughput $\texttt{RC}_L$, instantiates two parallel modules of $\texttt{RC}_A$, thus supporting a batch size of 26 and still remaining within an acceptable range for resource utilization.

\subsection{Extending \sys}

The modular and generalized design approach adopted in \sys naturally supports extensions to additional hash functions and other finite fields. To extend \sys to additional ZK-friendly hash functions, the following steps can be taken:

\noindent\textbf{1. Identify Key Arithmetic Primitives:}
The first step involves analyzing the new target hash functions to identify their core arithmetic primitives. Typically, these include modular multiplication, power mapping, inverse power mapping, quadratic evaluations, and lookups or decomposition/recomposition techniques. Thanks to \sys's modular architecture, existing modules can readily be adapted or extended to accommodate these new requirements.

\noindent\textbf{2. Modular Adaptation and Instantiation:}
Once the essential operations have been identified, the relevant existing modules within \sys can be instantiated or slightly modified. For example, the existing reconfigurable modular multiplier and fast division modules designed for RC can support any hash function employing similar decomposition and recomposition operations. Similarly, the fast power mapping module leveraging the square-and-multiply algorithm can accommodate exponentiation-based primitives in newly added hash functions.

\noindent\textbf{3. Field Generalization:}
Extending \sys to support different finite fields beyond BN254, such as the Goldilocks 64-bit field \cite{ernstberger2024zk}, involves minimal overhead, provided the arithmetic primitives remain similar. For prime fields with different sizes or characteristics, the underlying modular arithmetic modules (e.g., multiplication, division) must be parameterized accordingly. This parameterization involves adjusting the bitwidths and pre-computing constants that are tailored to the parameters of the new field. However, changing the field may necessitate repeated computations and validations to ensure that equivalent security levels are maintained compared to BN254 (or a similarly large field), especially when using smaller fields that might require additional rounds or operations to reach desired security guarantees.

\section{Results}

\subsection{Experimental Setup}
\sys's hardware modules are developed in Vivado 2024.2 with Verilog and synthesized on a Virtex Ultrascale+ device with part number \texttt{xcu250-figd2104-2L-e}. The CPU benchmarks we report are from those reported in \cite{grassi2022reinforced}, due to the extensive, and seemingly fair, evaluation that the authors conducted. We replicate these benchmarks on a 128GB RAM, AMD Ryzen 3990X CPU to ensure their accuracy and fairness with the \texttt{ZKFriendlyHashZoo} repository \cite{HashZKP}. All designs and benchmarks can be found in our open-source repository.
\nges{ we can also state that we reproduce the hash functions to validate their implementation?? I think Nojan said something like that in the meeting.}

\subsection{\sys Evaluation}
\nojan{need to say what the parameters are (e.g. num of rounds, rate, state size)}
\tk{We should say we measure the latency for permutations only. But we also created a top level wrapper for the design and synthesized it}

All three hashes are implemented over the BN254 field in this work. For benchmarking the hashing, we focus on measuring the performance of the permutation operation for software and hardware both. This is because these hashes use sponge construction, wherein hashing an input message boils down to repeated invocations of the permutation operation, interleaved with some state modification operations that have negligible runtime cost relative to the permutation. For simplicity, we simply measure the cost of one invocation of the permutation operation. We also ignore the overhead of I/O traffic in and out of the FPGA board, and assume that there is always a batch of requests ready such that the pipeline is fully saturated. Moreover, note that the throughput (and latency) figures in Tables \ref{tab:hashemall_benchmark} and \ref{tab:hash_performance} are amortized.

From Table~\ref{tab:hashemall_benchmark}, the trade-off between the $(\cdot)_A$ and $(\cdot)_L$ variants can be seen clearly. For such a multiplication-intense workload, the DSPs are the most valuable resource on the board. \texttt{Griffin}$_L$ achieves a throughput thrice as high as that of \texttt{Griffin}$_A$, but at the cost requiring thrice as much DSPs (and twice as much LUTs and power draw). Although we chose Virtex Ultrascale+ as our target board, \sys could in fact fit into a much smaller board based on the resource usage figure. For area-optimized design, we report a comparable resource usage figure to \cite{ahmed2024amaze}, making it suitable for edge applications.


We report \sys performance against CPU performance in Table ~\ref{tab:hash_performance}. For latency optimized designs, \sys outperforms the CPU implementation by at least an order of magnitude, with the smallest speedup being \texttt{Griffin}$_L$. Our best implementation, \texttt{RC}$_L$, outperforms CPU by 23.44$\times$, has a comparable run time with SHA-3, which has a reported latency of 419.2 ns in the same environment as the CPU benchmarks in Table ~\ref{tab:hash_performance} \cite{grassi2022reinforced}. This results also out This enables the use \sys's \texttt{RC}$_L$ as a practical candidate for hashing in ZKP applications built for proving systems that support lookups. In all scenarios, \sys provides a new state-of-the-art solution for practical development and deployment of ZKP applications in real-world systems.

\section{Conclusion}

This work presented \sys, the first realization of Griffin, Rescue-Prime, and Reinforced Concrete on reconfigurable hardware. \sys achieves state-of-the-art performance by first building highly-optimized modules for finite field arithmetic. These modules serve as the foundation of the end-to-end hash function accelerators. Through our experimental evaluation, we show that all the hash functions that are accelerated by \sys exhibit speedup of \textit{at least} an order of magnitude compared to optimized CPU solutions, alongside significant speedup of the Reinforced Concrete hash, achieving a comparable latency to SHA-3. The separate area-optimized and latency-optimized designs for \textit{each} hash function enables \sys to be a solution in many different settings, including resource-constrained settings. The open-source nature of our solution enables easy integration into any system, as well as presenting a modular approach that can be applied to several ZK-friendly hashes that operate over different finite fields.
\sys presents a paradigm shift in the realm of ZK-friendly hashing by making these hash functions practical for real-world adoption in zero-knowledge applications.

\section{Acknowledgments}
This work is supported by the U.S. Army/Department of Defense award number W911NF2020267 and the Defense Advanced Research Projects Agency (DARPA) Proofs program under Grant No. HR0011-23-1-0006.

\bibliographystyle{IEEEtran}
\bibliography{refs}

\end{document}